\documentclass[journal]{IEEEtran}

\usepackage{mathtools,amssymb,lipsum, nccmath}

\usepackage{cuted}
\usepackage{cite}

\usepackage{amsmath,amssymb,amsfonts}
\usepackage{mathtools}
\usepackage[linesnumbered,ruled,vlined]{algorithm2e}
\usepackage{algcompatible}

\SetCommentSty{mycommfont}
\usepackage[flushleft]{threeparttable}
\usepackage{footnote}
\usepackage{nicematrix,enumitem,booktabs}
\makesavenoteenv{tabular}
\usepackage{xcolor}
\usepackage{color}
\usepackage{url}
\usepackage{mdwmath}
\usepackage{mdwtab}
\usepackage[all]{xy}
\usepackage{array}
\usepackage[tight,footnotesize]{subfigure}
\usepackage{stfloats}
\usepackage{bbm} 
\usepackage{adjustbox}
\usepackage{orcidlink}
\usepackage{caption}
\usepackage{subcaption}
\usepackage{graphicx}
\DeclareGraphicsExtensions{.pdf,.jpeg,.eps}
\usepackage{multicol}
\usepackage{graphicx}
\usepackage{setspace}

\usepackage{ulem,bm}

\SetKwInput{KwData}{Input}
\SetKwInput{KwResult}{Output}

\begin{document}
	\title{Acousto-optic reconstruction of exterior sound field based on concentric circle sampling with circular harmonic expansion}
	
	\author{Phuc Duc~Nguyen\,\orcidlink{0000-0002-7136-8924},~\IEEEmembership{Member,~IEEE,}
		Kenji~Ishikawa\,\orcidlink{0000-0003-0423-566X},~\IEEEmembership{Member,~IEEE,},
		Noboru~Harada\,\orcidlink{0000-0002-1759-4533},~\IEEEmembership{Senior Member,~IEEE,}
		and~Takehiro~Moriya\,\orcidlink{0000-0003-4591-1273},~\IEEEmembership{Life Fellow,~IEEE}
		\thanks{The manuscript was accepted and published in IEEE Transactions on Instrumentation and Measurement, vol.\,74, June 2025, DOI: 10.1109/TIM.2025.3577852 (\emph{Corresponding author: Kenji Ishikawa, ke.ishikawa@ntt.com})}
		\thanks{P. D. Nguyen, K. Ishikawa, N. Harada, T. Moriya are with NTT Corporation. Atsugi R\&D Center. 3-1, Morinosato-Wakamiya, Atsugi, Kanagawa, 243-0198, Japan. (nguyenducphuc@ieee.org, ke.ishikawa@ntt.com, harada.noboru@ntt.com, takehiro.moriya@ntt.com)}
	}
	\markboth{}%
	{Shell \MakeLowercase{\textit{et al.}}: Bare Demo of IEEEtran.cls for IEEE Journals}
	
	\maketitle
	
	\begin{abstract}
		Acousto-optic sensing provides an alternative approach to traditional microphone arrays by shedding light on the interaction of light with an acoustic field. Sound field reconstruction is a fascinating and advanced technique used in acousto-optics sensing. Current challenges in sound-field reconstruction methods pertain to scenarios in which the sound source is located within the reconstruction area, known as the exterior problem. Existing reconstruction algorithms, primarily designed for interior scenarios, often exhibit suboptimal performance when applied to exterior cases. This paper introduces a novel technique for exterior sound-field reconstruction. The proposed method leverages concentric circle sampling and a two-dimensional exterior sound-field reconstruction approach based on circular harmonic extensions. To evaluate the efficacy of this approach, both numerical simulations and practical experiments are conducted. The results highlight the superior accuracy of the proposed method when compared to conventional reconstruction methods, all while utilizing a minimal amount of measured projection data.
	\end{abstract}
	
	\begin{IEEEkeywords}
		Acousto-optic sensing, sound field reconstruction, tomography, wave expansion.
	\end{IEEEkeywords}
	
	\IEEEpeerreviewmaketitle
	
\section{Introduction}
\IEEEPARstart{A}{}cousto-optics sensing methodologies utilize light as the principal sensing modality to non-intrusively sample acoustic fields within a given volume through a sparse array of optical measurements performed remotely. In contrast to conventional electromechanical transducers, acousto-optic methods do not yield localized acoustic pressure measurements due to the acousto-optic interaction taking place throughout the entire optical path \cite{torras2012sound}. 
Acousto-optic sensing is gaining attention due to its ability to provide non-contact imaging of sound fields, even at high frequencies or in the immediate vicinity of sound sources, a task typically challenging for traditional microphone measurements.
Among numerous acousto-optic sensing methods developed so far~\cite{AcousticsToday}, precision optical interferometry has played a central role in measuring audible sound, such as laser Doppler vibrometry (LDV) \cite{Nakamura2002, Harland2002,torras2012sound}, parallel phase-shifting interferometry and digital holography \cite{Ishikawa2016,Matoba2014,Ishikawa2018,Rajput2019,Takase2021,Rajput2021,Zhong2022}, and midfringe locked interferometry \cite{Ishikawa2021}. 

Computed tomography, frequently observed in medical imaging or nondestructive testing domains, enables the reconstruction of the spatial distribution of physical property from its projection measurements \cite{flannery1987three}. In acousto-optic sensing, the tomographic reconstruction of the sound field is essential to obtain sound pressure distributions from the measured laser projection data.

\begin{figure}[t]	
\centerline{\includegraphics[width=\columnwidth]{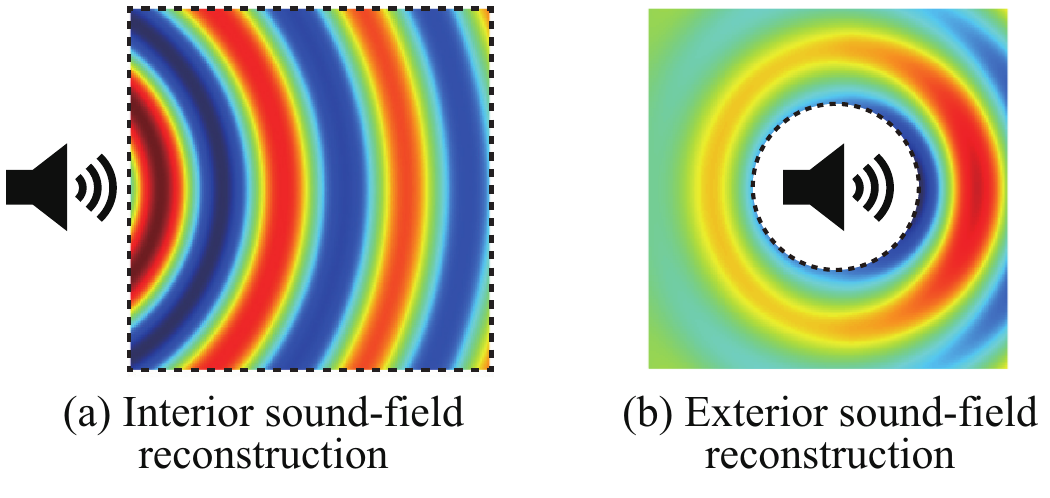}}
	\caption{Illustration of (a) interior and (b) exterior sound-field reconstruction problems.}
	\label{Fig_1}
\end{figure}

A challenge in the existing sound-field reconstruction methods is the so-called {\it exterior problem} where the sound source is located inside the reconstruction area. 
Fig.~\ref{Fig_1} illustrates the two sound-field reconstruction problems. The {\it interior problem}\textcolor{black}{, defined as a situation in which no sound source exists within the measurement domain~\cite{Koyama2024},} reconstructs the sound field inside a closed space (represented by a dashed line) where a sound source is located outside the space. The {\it exterior problem}\textcolor{black}{, defined as a situation in which a sound source located inside the measurement domain~\cite{Ochmann1999},} reconstructs the sound \textcolor{black}{propagating outwards from the source}.
In physical acoustics, the exterior and interior sound field problems are governed by different mathematical formulations~\cite{williams2000fourier}: the latter is formulated using radial functions that take finite values at the origin, while the former employs radial functions that represent outward-propagating waves; therefore, model-based approaches tailored to interior problems are not directly applicable to exterior problems, and vice versa (see Appendix for a detailed discussion).

Since the existing reconstruction algorithms are optimized for the interior reconstruction problem~\cite{yatabe2017acousto, verburg2021acousto, ishikawa2021physical, Verburg2022}, they exhibit suboptimal performance for the exterior situation. Nevertheless, the reconstruction of the external sound field is a critical problem that includes essential applications such as directivity measurement of sound sources. Directivity measurements are essential for various purposes, such as optimizing the design of acoustic devices, evaluating the performance of sound systems, and assessing the impact of noise pollution. Moreover, solving exterior problems has an important effect on estimating the radiated sound field to an infinite region. This estimation is vital for applications where the sound propagates over long distances, such as outdoor environments, large industrial facilities, or open spaces. 

In this study, we present a novel exterior reconstruction method founded upon a concentric circle sampling scheme, utilizing circular harmonics as the expansion coefficients. To the best of our knowledge, this research represents the pioneering effort to facilitate the acousto-optic reconstruction of exterior sound fields, even in cases where the sound source is situated within the confines of the reconstruction area. 

The rest of this paper is structured as follows: Section II summarizes related works and drawbacks of current sound field reconstruction approaches in solving the interior problem. Section III introduces the proposed method, which is comprised of a concentric circle sampling technique and circular harmonic expansion of the sound field. Section IV presents the simulation setups and the corresponding simulated results. This is followed by the exposition of the experiment setups and the actual experimental results. Section V then meticulously presents and analyzes the experimental findings. Section VI engages in relevant discussions to shed light on the implications of the results. Finally, Section VII encapsulates the research with a conclusive summary of key findings and their broader significance.

\section{Related works}
The filtered back-projection (FBP) method is the most widely used tomographic reconstruction technique \cite{kak2001principles}. While this method is straightforward and commonly applied \cite{Oikawa2005, torras2012sound, koponen2019acoustic, HERMAWANTO2020}, it may not yield optimal accuracy when reconstructing sound fields.
In cases where uniform scanning of the object cannot be achieved, algebraic reconstruction techniques (ARTs) are often utilized \cite{kak2001principles}. These methods approach the problem by representing it as a linear system with unknown variables and equations.
Additionally, iterative algorithms, such as the simultaneous algebraic reconstruction technique (SART) and iterative reconstruction techniques (IRT), have been proposed in the literature \cite{andersen1984simultaneous, herman2009fundamentals, natterer2001mathematics}. However, these algorithms are not specifically designed for sound field reconstruction, as they do not impose constraints related to the characteristics of the sound field. In contrast, previous studies have demonstrated that physical model-based sound-field reconstruction methods, which explicitly account for the physical properties of sound fields, significantly outperform conventional methods such as FBP, ARTs, SART, and IRT \cite{yatabe2017acousto, verburg2021acousto, ishikawa2021physical, Verburg2022}. 

A typical reconstruction algorithm in the physical model-based group is the plane-wave expansion (PWE) \cite{verburg2021acousto, Verburg2022}. PWE models the sound field by the summations of elementary plane waves, which is derived from the Helmholtz equation in Cartesian coordinates. PWE assumes that no object or scatterer exists inside the reconstruction region, which enforces that the field to be reconstructed must be the interior problem. If the PWE is used for the exterior problems, it produces significant reconstruction errors due to the violation of the physical model.
Therefore, in this study, we introduce a new sound field reconstruction algorithm based on circular harmonic expansion to solve the exterior problem. To the best of our knowledge, our proposed method is the first to solve the exterior reconstruction problem when the sound source is located within the reproduction area. Furthermore, by introducing this method, we anticipate a reduction in algorithmic complexity by using fewer projections compared to traditional sound field reconstruction methods.

\section{Proposed method}
\subsection{Principle of acousto-optic sensing}

\begin{figure}[!t]	
\centerline{\includegraphics[width=\columnwidth]{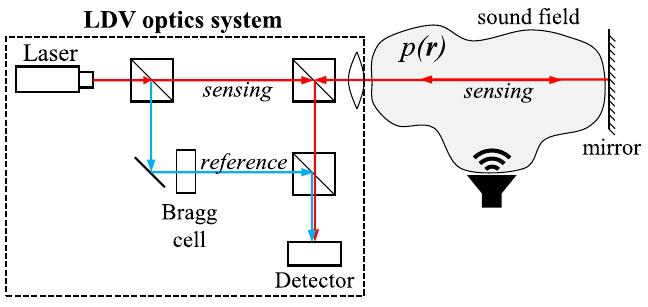}}
	\caption{Acousto-optic sensing utilizing a Laser Doppler Vibrometer (LDV) device.}
	\label{Fig_2}
\end{figure}

Fig.~\ref{Fig_2} presents an exemplification of the LDV, which serves to gauge acoustically induced phase shifts of light. The functioning of this apparatus involves the generation of a laser beam, subsequently bifurcated into two distinct branches: the sensing branch and the reference branch. The sensing beam is then directed through the sound field $p(\bm{r})$, wherein it undergoes backscattering upon interacting with the mirror surface and is eventually retrieved at the LDV optics system. The sensing beam experiences the phase modulation caused by sound due to the acousto-optic effect, which is detected by the LDV.
The sound field denoted as $p(\bm{r})$ represents the sound pressure at the position $\bm{r}$ in space. In this paper, since our reconstruction algorithm is defined in the frequency domain, we define the sound field as in the frequency domain with the harmonic oscillation $\omega$. Therefore, $p(\bm{r}, \omega)$ represents the complex amplitude at $\omega$, which is derived using the Fourier transform to the measured temporal signal. Hereafter, for simplicity, we denote $p(\bm{r}, \omega)$ as $p(\bm{r})$ by omitting $\omega$.

A notable characteristic of the acousto-optic sensing is that the detected signal is proportional to the line integral of sound pressure along the sensing beam~\cite{torras2012sound}.
The underlying physical principle involves the phase alteration of a laser beam as it passes through the sound field, wherein the phase shift of light is determined as
\begin{equation}
\label{eq0}
\textcolor{black}{\delta\varphi=\frac{2\pi}{\lambda_l}j\omega\int_{L}{p(\bm{r})dl}},
\end{equation}
where $j = \sqrt {- 1}$, $\bm{r}$ is the position vector, $l$ serves as the integration variable along the laser beam, $L$ represents the length of the laser path, and $\lambda_l$ is the optical wavelength. Equation (\ref{eq0}) shows that measured phase shifts are proportional to the sound field pressure along the laser beam $\int_{L}{p(\bm{r})dl}$.

\subsection{Concentric circular sampling}
\begin{figure}[t]	
\centerline{\includegraphics[width=\columnwidth]{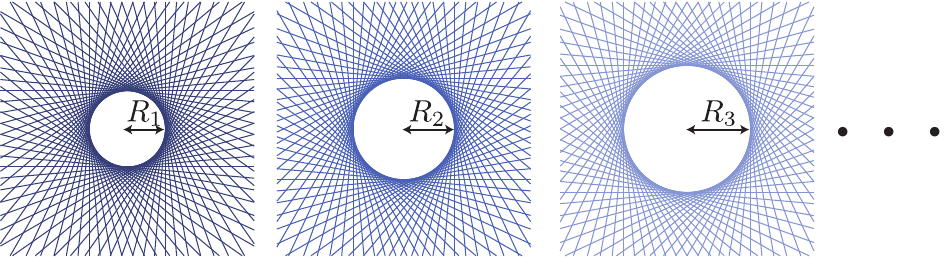}}
	\caption{The proposed concentric circle sampling samples a sound field along the tangent line of a circle of radius $R_i$ at given angular intervals. The lines in the images represent measurement laser trajectories when the angular interval is 5$^\circ$.}
	\label{Fig_3}
\end{figure}

For exterior reconstruction problems, since a sound source should be located in the center part of the reconstruction area (as depicted in Fig.~\ref{Fig_1}(b)), sensing laser paths must avoid this part. Considering practical experimental implementation, we propose the concentric circle sampling scheme as shown in Fig.~\ref{Fig_3}. The projections $S$ are sampled by laser beams surrounding the circle with radius $R$ and rotation $\phi$. This sampling scheme is easily achieved by rotating the sound source while keeping the laser beam fixed. In order to enhance the fidelity of reconstructed sound fields, it is possible to augment the acquisition of projections by incorporating numerous circles with varying radii, such as $R_1$, $R_2$, and $R_3$ as depicted in Fig.~\ref{Fig_3}, or by employing smaller angular rotation increments. In this context, Fig.~\ref{Fig_3} illustrates 72 projections derived from a single concentric circle when the angular interval is 5$^\circ$. This choice of the number of projections was made to strike a judicious balance between the reconstruction quality and the computational complexity inherent in the reconstruction process. Another important factor is the optimal number of concentric circles to use in sound field reconstruction. In an upcoming section, we will share initial findings and discuss this issue.

\subsection{Circular harmonic expansion of sound field}
\label{CHE_subsection}
In this section, we expound upon the sound field reproduction technique employing circular harmonics, focusing on the global exterior sound field as the area of interest. The acoustic field generated by an arbitrary sound source within a two-dimensional exterior space can be accurately depicted through the application of circular harmonic expansion (CHE), as exemplified in the works \cite{williams2000fourier, wang2015optical, arsenault1986properties}. This representation can be expressed as follows (see Appendix):
\begin{equation}
\label{eq1}
p(\bm{r}) = p(r,\phi) = \sum\limits_{n =  - \infty }^\infty  {{a_n(\omega)}} {H_n^{(2)}}(kr){e^{jn\phi }},
\end{equation}
where $(r, \phi)$ are the radial distance and angle in the polar coordinate, ${a_n(\omega)}$ represents the sound field coefficient for the order $n$, and ${H_n^{(2)}}\left(  \cdot  \right)$ is the Hankel function of the second kind.
Note that the scalar $r$ differs from the position vector $\bm{r}$.
The symbol $k$ corresponds to the wave number of sound, which is given by $k = \omega/c$, where $c$ denotes the propagation velocity.

In our reconstruction method, the parameter $n$ was chosen flexibly based on the sound signal's frequency and the smallest radius of the concentric circle. This choice was important to accurately represent the reconstructed sound field.


\subsection{Reconstruction algorithm}
According to the acousto-optic theorem and the circular harmonic expansion of the sound field, the projection $s_m$ can be written as
\begin{equation}
\label{eq2}
{s_m} = \int {p(r,\phi)d{l_m} \approx \sum\limits_{n =  - N}^N {{a_n}\int {{H_n^{(2)}}(kr){e^{jn\phi }}d{l_m}} } },
\end{equation}
where $l_m$ represents the sensing laser trajectory of the $m$th measurement, and the infinite sum of circular harmonics is truncated by order $N$.
Equation (\ref{eq2}) can be expressed algebraically as follows.
\begin{equation}
\label{eq3}
\mathbf{s} \approx \mathbf{H}\mathbf{a},
\end{equation}
where $\mathbf{H} \in {\mathbb{C}^{M \times (2N+1)}}$ is a matrix with elements ${h_{m,n}} = \int {{H_n^{(2)}}(kr){\mathrm{e}^{jn\phi}}d{l_m}}$. $\mathbf{a} \in {\mathbb{C}^N}$ are the expansion coefficients of the $N$ circular harmonic waves in the expansion. These coefficients can be determined through the regularized least squares method.
\begin{equation}
\mathbf{\tilde a} = \arg \mathop {\min }\limits_{\mathbf{a} \in {\mathbb{C}^N}} {\left\| \mathbf{a} \right\|_2}
\end{equation}
subject to ${\left\| {\mathbf{s} - \mathbf{H}\mathbf{a}} \right\|_2} \le \varepsilon$, where $\varepsilon$ is the permitted discrepancy between measurement $\mathbf{s}$ and $\mathbf{H}\mathbf{a}$. \textcolor{black}{The coefficient vector 
$\mathbf{a}$ is independently estimated for each frequency and is obtained using Tikhonov regularization.}

After obtaining the expansion coefficients $\mathbf{\tilde a}$, the reconstructed sound field is calculated as
\begin{equation}
\mathbf{\tilde p} = \mathbf{G}\mathbf{\tilde a}
\end{equation}
where $\mathbf{\tilde p} \in {\mathbb{C}^L}$ denote estimated pressure at positions $\bm{r}_1, ..., \bm{r}_L$, and $\mathbf{G} \in {\mathbb{C}^{L \times (2N+1)}}$ includes elements ${g_{l,n}} = {H_n^{(2)}}(kr_l) {\mathrm{e}^{jn\phi_l}}$.

\section{Simulation}
To verify the validity of the proposed method, a 2D sound field was generated, and the projection data (line integrals of the sound field) were obtained numerically. These data were used as input for both the proposed method and the comparative methods to evaluate reconstruction accuracy.

\subsection{Reference multi-source complex sound field}
\label{sectionIII_A}
As reference sound fields for simulating the exterior reconstruction problem, we generated sound fields within the area of 0.7$\,$m $\times$ 0.7$\,$m by the superposition of five point sources located within a circular area of minimum radius $R_\mathrm{min} = 0.3\,$m. 
The generated sound field, composed of five point sources, is defined as follows:
\begin{equation}
    \label{eq_soundfield_gen}
    p(\bm{r}) = A \sum^{5}_{i=1} \frac{1}{|\bm{r} - {\bm{r}_{0,i}}|} \mathrm{e}^{j (k |\bm{r} - {\bm{r}_{0,i}}| + \theta_{0,i})},
\end{equation}
where $A$ is the amplitude of the sound field, $\bm{r}_{0,i}$ and $\theta_{0,i}$ are the position and phase of $i$th point source, respectively, and $\left| \bm{r} - {\bm{r}_{0,i}} \right|$ denotes the distance between $\bm{r}$ and $\bm{r}_{0,i}$. For reference sound-field generation, we calculated in Cartesian coordinate, i.e., $\bm{r} = (x, y) = (r \cos \phi, r \sin \phi)$. We chose the positions and phase of the five point sources as $\bm{r}_{0,i} = (x_{0,i}, y_{0,i}) = \{(0,0), (-0.05,0), (0, -0.05), (0.05, 0), (0, 0.05)\}$ and $\theta_{0,i} = \{\pi/6, 2 \pi/6, 3\pi/6, 4\pi/6, 5\pi/6\}$ for $i = \{1,2,3,4,5\}$. The units for x and y are meters.

\begin{figure*}[!t]
\centering
\includegraphics[width=0.95\textwidth]{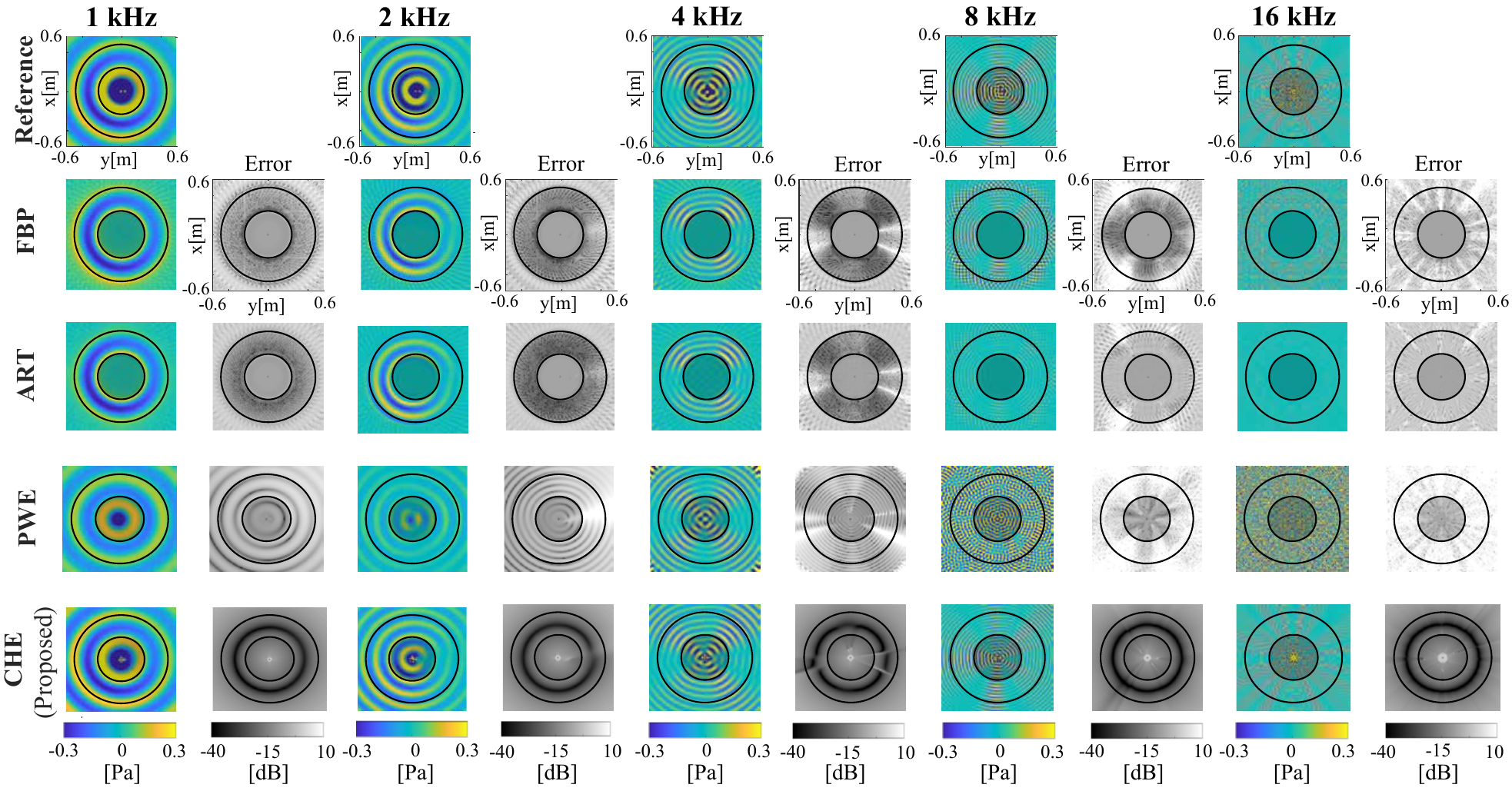}
\caption{A comparative analysis was conducted on sound fields reconstructed by different reconstruction algorithms, namely FBP, ART, PWE, and the proposed CHE. The reference sound fields were constructed using multiple sound point sources distributed in the proximity of a central location and across a range of frequencies.}
\label{Fig_5}
\end{figure*}

\begin{figure*}[!t]
\centering
\includegraphics[width=0.95\textwidth]{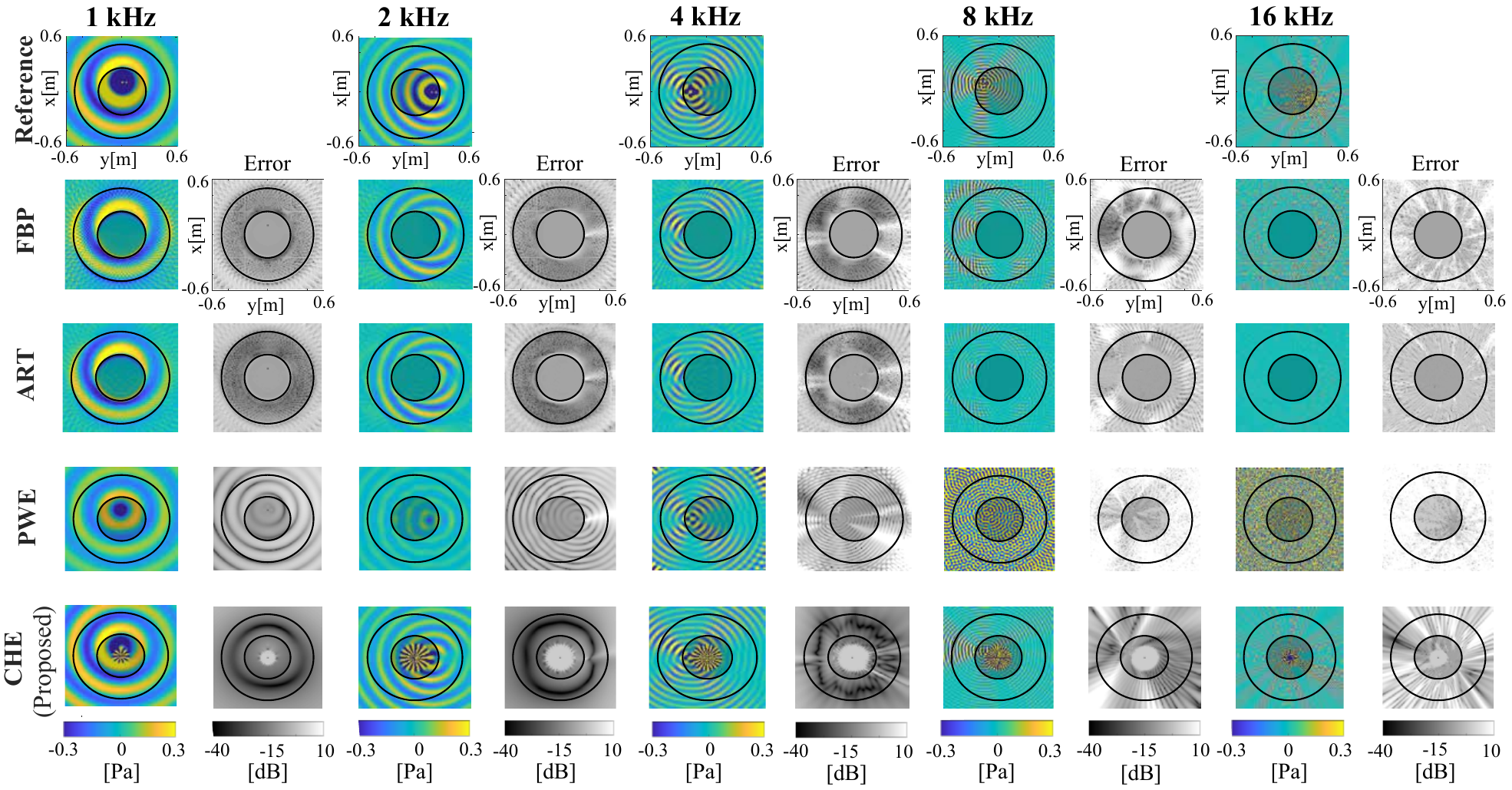}
\caption{A comparative evaluation was conducted to assess the sound field reconstruction quality of various reconstruction algorithms, specifically FBP, ART, PWE, and the proposed CHE. The reference sound fields were meticulously composed, featuring multiple sound point sources positioned at locations distanced from the central point. The specified locations for the groups of sound sources were as follows: at 1$\,$kHz: (0, 0.12), 2$\,$kHz: (0.2, 0), 4$\,$kHz: (-0.2, 0), 8$\,$kHz: (-0.2, 0.1), and 16$\,$kHz: (0.2, -0.1). These sound fields were generated at distinct frequencies for the express purpose of evaluation.}
\label{Fig_6}
\end{figure*}

To enable a rigorous comparative analysis, we conducted simulations involving sound fields generated at different frequencies 1$\,$kHz, 2$\,$kHz, 4$\,$kHz, 8$\,$kHz, and 16$\,$kHz. For each frequency, we assessed two distinct scenarios wherein sound sources were positioned both at the center and off-center locations. 
The generated centered and off-centered sound fields are displayed at the tops of Figs.~\ref{Fig_5} and ~\ref{Fig_6}, respectively. Note that the off-centered sound fields are simply centered sound fields shifted by a certain amount. This enables the evaluation of reconstruction performance when the origin of the sound field does not align with the origin of the concentric circular sampling. For our simulations, we numerically calculated the projection $s_m$ from the reference sound fields.

\subsection{Simulated sound-field reconstruction}
\label{simulatedreconstruction}
In order to assess the quality of the reconstructions, we used reconstruction error \cite{ren2020two}, which is calculated using $\varepsilon (\bm{r}) = 10{\log _{10}}\left\{ {{{{{{{\left| {\tilde p(\bm{r}) - p(\bm{r})} \right|}^2}} \mathord{\left/{\vphantom {{{{\left| {\tilde p(\bm{r}) - p(\bm{r})} \right|}^2}} {\left| {p(\bm{r})} \right|}}} \right.
\kern-\nulldelimiterspace} {\left| {p(\bm{r})} \right|}}}^2}} \right\}$, where $\tilde p(\bm{r})$ and $p(\bm{r})$ are the reconstructed and reference sound pressure at position $\bm{r}$, respectively. We computed the reconstruction error on a pixel-wise basis and juxtaposed the resulting error images alongside each corresponding reconstructed sound field solely for evaluative purposes. \textcolor{black}{The error or difference, denoted as $\varepsilon (\bm{r})$, represents the discrepancy between the reconstructed sound field and the original sound field, and is expressed in decibels (dB). Therefore, the lower the $\varepsilon (\bm{r})$ value (shown as dark black regions in the comparison images in Figs.~\ref{Fig_5} and ~\ref{Fig_6}), the better the performance of the sound field reconstruction algorithm. Conversely, higher $\varepsilon (\bm{r})$ values (represented by white regions in Figs.~\ref{Fig_5} and ~\ref{Fig_6}) indicate poorer performance of the reconstruction algorithms}.

As discussed in Section II, various techniques are currently employed for interior sound field reconstruction. From the most widely used categories, we selected three representative algorithms: FBP, which exemplifies the group utilizing uniform projection structures; ART, which represents the group employing non-uniform projection structures; and PWE, which characterizes the group leveraging wave expansion properties. As for the evaluation criteria, we have chosen to evaluate the reconstructed sound field through visualization, calculate the reconstruction errors compared to the original sound field, and use the most common metric, normalized mean square error (NMSE), for cases where the sound source is at a central (center) or off-center position, as well as for both cases with and without Gaussian noise in the sound field. To ensure fair comparisons, we maintained a uniform number of projections (M$\,$=$\,$2232) across different reconstruction algorithms: FBP, ART, PWE, and the proposed CHE-based approach. In the case of the PWE method, our simulations entailed the use of 200 plane waves. In the exterior sound field reconstruction problem, the sound source is designated to be positioned at the center of the sound field. We define the distance from the sound source to the laser path as $R$, where $R_\mathrm{min}\leq$ $R \leq R_\mathrm{max}$. In the simulation and experimental result images presented in this article, two circles with radii $R_\mathrm{min}$ = 0.3\,m and $R_\mathrm{max}$ = 0.6\,m are consistently depicted. To assess the efficacy of sound field reconstruction through visualization, we are only concerned with the sound field region within the boundaries defined by two circles $R_\mathrm{min}$ and $R_\mathrm{max}$.
 
Figure~\ref{Fig_5} depicts the reconstructed sound fields and reconstruction errors of the four methods: FBP, ART, PWE, and CHE. These reconstructions were performed with the sound sources located at the central coordinates (0, 0). We employed $N\,$=$\,10, 15, 20, 30, 40$ for CHE for frequencies: 1$\,$kHz, 2$\,$kHz, 4$\,$kHz, 8$\,$kHz, and 16$\,$kHz, respectively.  
The results show that both FBP and ART successfully reconstructed sound fields that generally match the reference field within the inner region between the two circles, $R_\mathrm{min}$ and $R_\mathrm{max}$. However, the reconstruction becomes entirely inaccurate outside this region, with the appearance of unwanted patterns at a resolution much finer than the wavelength, which is unrelated to the actual sound field.
For PWE, the reconstruction errors are apparently higher than those of FBP and ART, indicating that the physical model employed for the interior problem is inappropriate for solving the exterior problem. 
In contrast, the proposed CHE approach consistently yielded the highest quality reconstructions among the four methods under consideration. The results also indicate that as the sound frequency increases, such as at 8\,kHz and 16\,kHz, conventional algorithms like FBP, ART, and PWE almost fail to reconstruct the sound field properly. In contrast, CHE successfully addresses the challenge of sound field reconstruction at high frequencies while maintaining a good level of reconstruction quality.

Figure~\ref{Fig_6} illustrates the reconstruction results in scenarios where sound sources are positioned at varying distances from the central location (off-center). For ease of implementation and to ensure a fair comparison, we used the same number of circular harmonics as in Fig.~\ref{Fig_5}, even though the off-center sound fields exhibit greater complexity. 
The outcomes reveal that FBP, ART, and PWE exhibit suboptimal reconstruction performances similar to the centered sound field scenario in Fig.~\ref{Fig_5}.
CHE delivers sufficient reconstruction quality for 1, 2, and 4 kHz, whereas it failed to reconstruct the field for higher frequencies at 8 and 16 kHz.

 \begin{figure}[!t]
\centering
\includegraphics[width=0.45\textwidth]{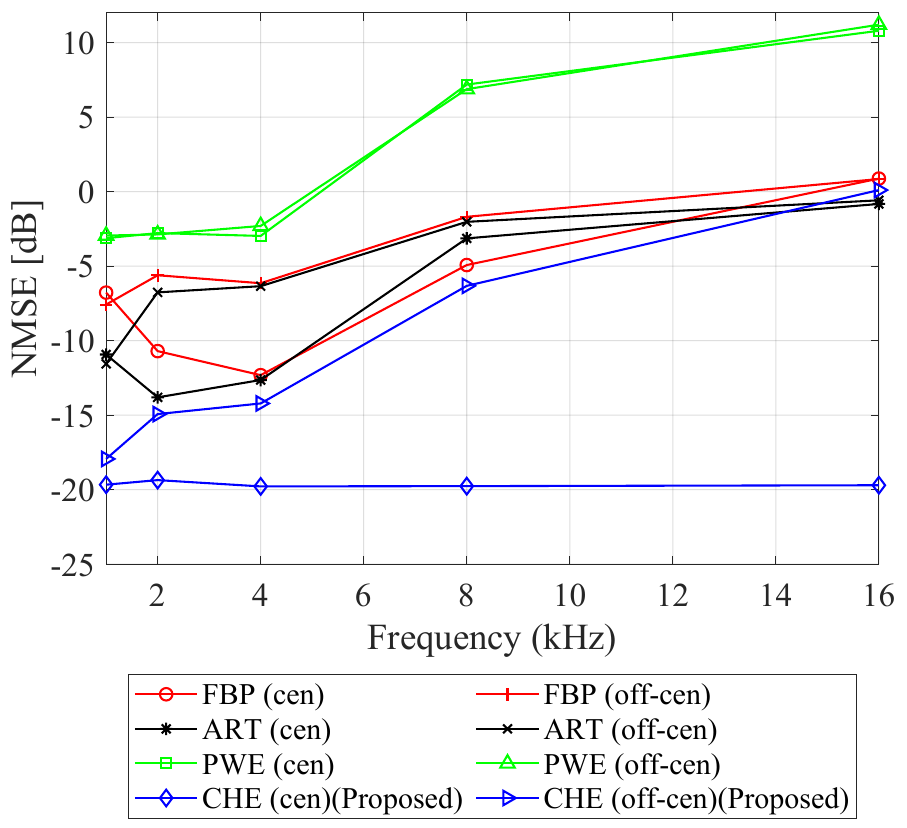}
\caption{NMSE evaluation of sound field reconstruction error using FBP, ART, PWE, and proposed CHE algorithms at different frequencies ranging from 1\,kHz to 16\,kHz. The sound sources for the reference sound field were positioned at both the center and off-center locations.}
\label{Fig_6_new}
\end{figure}


\begin{figure}[!t]
  \centering
  \subfigure[]{\label{Fig_7a}\includegraphics[width=0.45\textwidth]{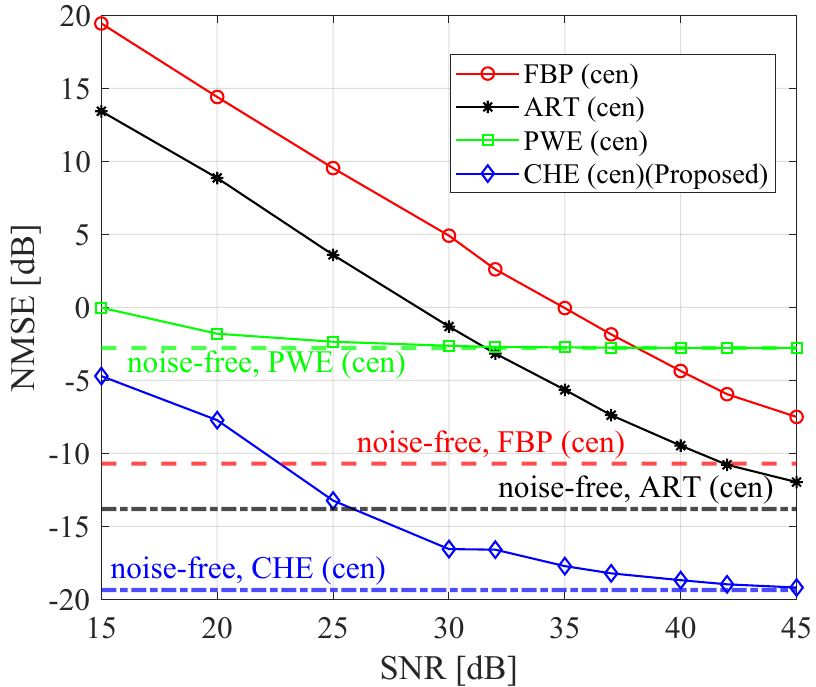}}\qquad
  \subfigure[]{\label{Fig_7b}\includegraphics[width=0.45\textwidth]{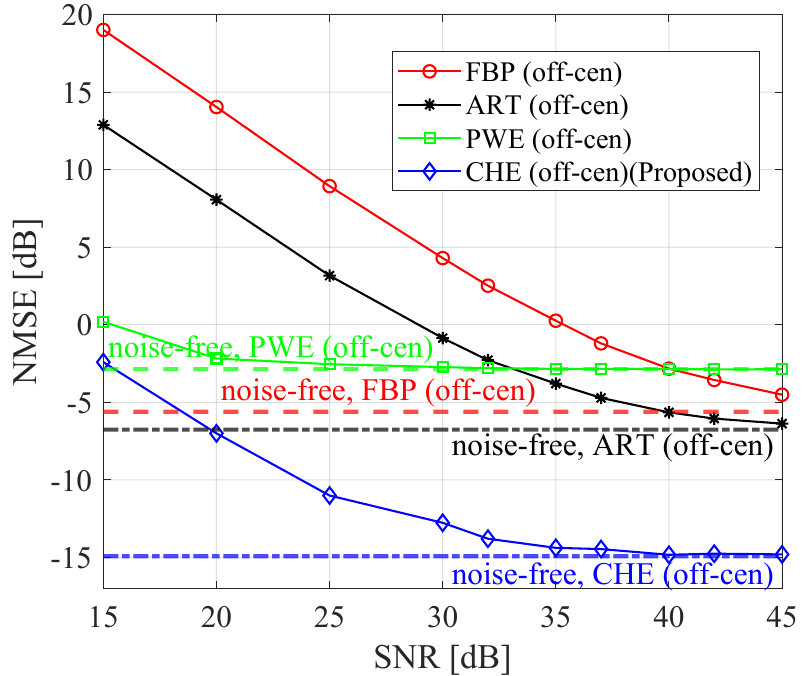}}
\caption{An NMSE evaluation was conducted to assess the quality of 2\,kHz sound-field reconstruction \textcolor{black}{in two cases: (a) when the sound source is at the center position (cen), and (b) when it is off-center (off-cen),} achieved by four distinct methods: FBP, ART, PWE, and the proposed CHE. This evaluation considered the influence of Gaussian noise on the reference sound field at varying signal-to-noise ratio (SNR) levels. Additionally, reconstructed sound fields were provided for reference when the reference sound field was in a noise-free environment.}
\label{Fig_7}
\end{figure}

For comparing the overall errors across the entire reconstruction area, the following NMSE in decibels (dB) is used; $\mathrm{NMS{E_{dB}}} = 20{\log _{10}}({{{{\left\| {\tilde p -  p} \right\|}_2}} \mathord{\left/
 {\vphantom {{{{\left\| {\tilde p - p} \right\|}_2}} {{{\left\| p \right\|}_2}}}} \right.
 \kern-\nulldelimiterspace} {{{\left\| p \right\|}_2}}})$.
Figure~\ref{Fig_6_new} shows the results.
It can be seen that the proposed CHE method exhibits the lowest NMSE for most conditions. Specifically, the reconstruction error is significantly lower when the sound field is centered compared to when it is off-center. This indicates that when applying the proposed method, it is important to align the center of the sound source with the center of the sampling as closely as possible. 
Among the conventional algorithms, FBP and ART exhibit similar performance for exterior sound field reconstruction, whereas PWE consistently performs the worst among the algorithms we tested. Therefore, it can be observed that the proposed CHE algorithm consistently outperforms current algorithms in the problem of exterior reconstruction for sound fields at different frequencies, whether the point source is located at the center or off-center.

To further assess reconstruction performance under varying noise conditions, different levels of white Gaussian noise were introduced to the 2$\,$kHz reference sound field. The NMSE performance of the four methods in reconstructing the noisy sound field is presented in Fig.~\ref{Fig_7}. \textcolor{black}{We evaluate the performance of the proposed solution when the sound source is placed at the center position (Fig.~\ref{Fig_7a}) and when it is slightly off-center (Fig.~\ref{Fig_7b}).}
The findings illustrate the persistent supremacy of the proposed CHE algorithm in comparison to the FBP, ART, and PWE across different noise levels. In high-noise regions, such as with an SNR between 15 and 30, it can be observed that the NMSE performance of the FBP and ART algorithms is poor, consistently remaining above 0. In contrast, the PWE algorithm performs better, occasionally reaching a threshold of -2$\,$dB. Meanwhile, the proposed CHE algorithm achieves excellent performance, with NMSE values ranging from -17$\,$dB to -5$\,$dB, even in the presence of strong noise with an SNR between 15 and 30. Moreover, in a noise-free environment, the sound fields reconstructed by the FBP, ART, and PWE methods reach saturation points of NMSE at -10.7$\,$dB, -13.8$\,$dB, and -2.78$\,$dB, respectively. In contrast, the sound fields reconstructed by the proposed CHE algorithm consistently exhibit ongoing quality improvement as SNR increases, approaching sound field quality levels obtained in a noise-free environment, with an NMSE of -19.4$\,$dB.

The reasons why the CHE method outperformed the other methods can be considered as follows. 
Since FBP and ART do not consider the characteristics of sound, the reconstruction is based purely on the data. This can lead to the reconstruction of non-acoustic components present due to the discrete observations or noise. As demonstrated by the reconstruction results at 1$\,$kHz and 2$\,$kHz, this approach permits the inclusion of unwanted patterns that are clearly different from the actual physical sound. This limitation should be the reason why FBP and ART perform worse than CHE.
PWE uses a superposition of plane waves coming from outside infinity to model a sound field in which no sound sources or objects exist inside the restored region. This is clearly different from the sound field used in this simulation, and the large restoration error is thought to be caused by the mismatch between the assumptions made by such a physical model and the data. On the contrary, CHE is constructed based on a physical model of a sound field radiated from a sound source inside of the smaller circle to the outside, so the assumptions of the algorithm and the data are consistent, and it has high restoration accuracy.
The simulation results confirmed the effectiveness of the proposed method for the external sound field reconstruction problems.

\begin{figure}[tb]
  \centering
  \subfigure[]{\label{Fig_8a}\includegraphics[width=0.4\textwidth]{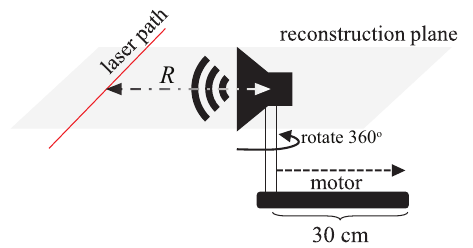}}\qquad
  \subfigure[]{\label{Fig_8b}\includegraphics[width=0.45\textwidth]{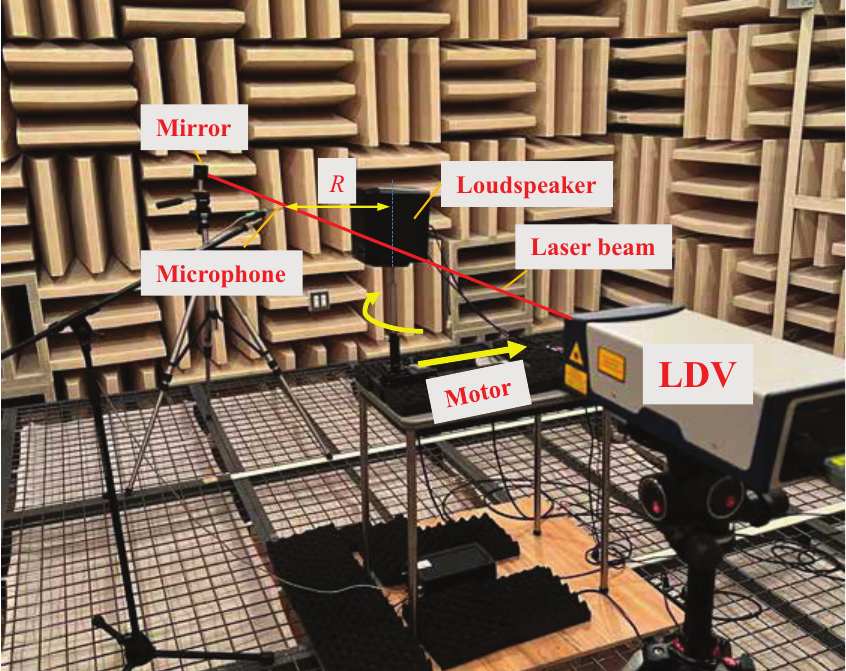}}
\caption{(a) Experimental setup for the concentric circle sampling scheme involved the rotation of the loudspeaker with a step size denoted as $\delta_\theta$, while the radius $R$ was adjusted using a unidirectional movement motor; (b) A photograph depicting our experimental system within the confines of an anechoic chamber.}
\label{Fig_8}
\end{figure}

\begin{table}[]
\caption{Experiment settings}
\resizebox{0.43\textwidth}{!}{
\begin{minipage}{0.5\textwidth}
\begin{center}
\begin{tabular}{|c|c|c|}
\hline
                              & \textbf{Parameter} & \textbf{Value} \\ \hline
Microphone      &  Type     & Br\"uel Kj\ae r 4939-A-011  \\ \hline                              
{}{}{Microphone amplifier} & Type      & Br\"uel Kj\ae r NEXUS  \\ \cline{2-3}       
                            & Signal delay        & $\approx10\,\mu$s    \\ \hline
Loudspeaker       &   Type    & Yamaha MS101III    \\ \hline
Stage controller & Type       &  OptoSigma SHOT-302GS   \\ \cline{2-3}
                              & Movement step            & 10$\,$mm         \\ \cline{2-3} 
                              & Rotation step            & 5$^\circ$         \\  \hline
{}{}{LDV} & Type     & Polytec VibroFlex         \\ \cline{2-3} 
                              & Decoder            & Displacement         \\ \cline{2-3} 
                              & Bandwidth          & 20$\,$kHz         \\ \cline{2-3} 
                              & Sensitivity        & 5 nm/V @M$\Omega$   \\ \cline{2-3} 
                              & Range              & 10$\,$nm          \\ \cline{2-3} 
                              & Signal delay       & 900$\,\mu$s         \\ \hline
{}{}{Anechoic chamber}   & Temperature      & 16.9$^\circ$C    \\ \cline{2-3} 
                                 & Humidity        & 28.4\,\%\\ \cline{2-3} 
                                  & Pressure        & 1000.4\,hPa \\ \hline
\end{tabular}
\label{tab1}
\end{center}
\end{minipage}
}
\end{table}

\section{Experiments}
We have confirmed through numerical simulations that the proposed method outperforms conventional methods in external sound field reconstruction. In this section, we will verify the effectiveness of the proposed method on real data by comparing the reconstructed sound fields from measurements taken in an anechoic chamber with measurement data obtained using microphones. To minimize noise that could affect the experimental results, we placed sound-absorbing materials to reduce environmental sound reflections to the lowest possible level.

\subsection{Experiment setup}
\label{experimentsetup}
Fig.~\ref{Fig_8} shows our experimental setup in an anechoic chamber. 
The sound field was generated by a loudspeaker positioned laterally in relation to the LDV. Instead of adjusting the position and angle of the laser path, we rotated and moved the loudspeaker while keeping the laser beam fixed in place to achieve the concentric circle sampling. Our method of scanning by moving and rotating the sound source limits the applicable sound sources. For example, it is difficult to apply it to heavy and large sound sources, human-played instruments, and sound sources that cannot produce the same sound repeatedly. On the other hand, it can be used effectively for loudspeakers, which are our main interest.
A mirror was positioned at a distance of 289$\,$cm from the laser window of the LDV. To adjust the distance from the center of the loudspeaker to the laser path $R$, we utilized a motor under the control of a LabVIEW program to manage both the rotation of the loudspeaker and its one-dimensional movement. The distance $R$ was incremented by 10$\,$mm for each step, ranging from $R_\mathrm{min}\,$=$\,0.3\,$m to $R_\mathrm{max}\,$=$\,0.6\,$m. 
The rotation step $\delta_{\theta}$ was fixed at 5$^\circ$, and a total of 72 projections were obtained for each circle sampling.
Therefore, the total number of the measurement $M = 31 \times 72 = 2232$.
The comprehensive information regarding the equipment and experimental parameters are listed in Table \ref{tab1}.

\textcolor{black}{The total measuring time for sampling one sound field includes several components. First, the sound source undergoes 72 rotations, with each rotation incremented by 5 degrees. For each rotation angle, 31 distance displacements are performed, moving the laser from the center of the sound source at a radius ranging from 0.3\,m to 0.6\,m in 1\,cm steps. Each displacement takes approximately 1 second for the mechanical bar movement and an additional 2\,s for data sampling. Furthermore, after each set of 31 displacements per rotation angle, there is a time loss of about 5\,s required to reset the laser position from R = 0.6\,m back to $R$ = 0.3\,m. Taking all these factors into account, the estimated total time required to complete the measurement process for one sound field is approximately 6,701\,s, equivalent to 1 hour, 51 minutes, and 41 seconds.}
\textcolor{black}{The time required for sound field reconstruction depends on the specifications of the test computer. On a Lenovo ThinkPad P52s equipped with an 8th Gen Intel Core i7 processor, NVIDIA Quadro P500 graphics card, and 32\,GB of DDR4 RAM, the reconstruction process takes less than one minute.}

\subsection{Harmonic sould-field reconstruction}

\begin{figure*}[!t]
\centering
\includegraphics[width=0.9\textwidth]{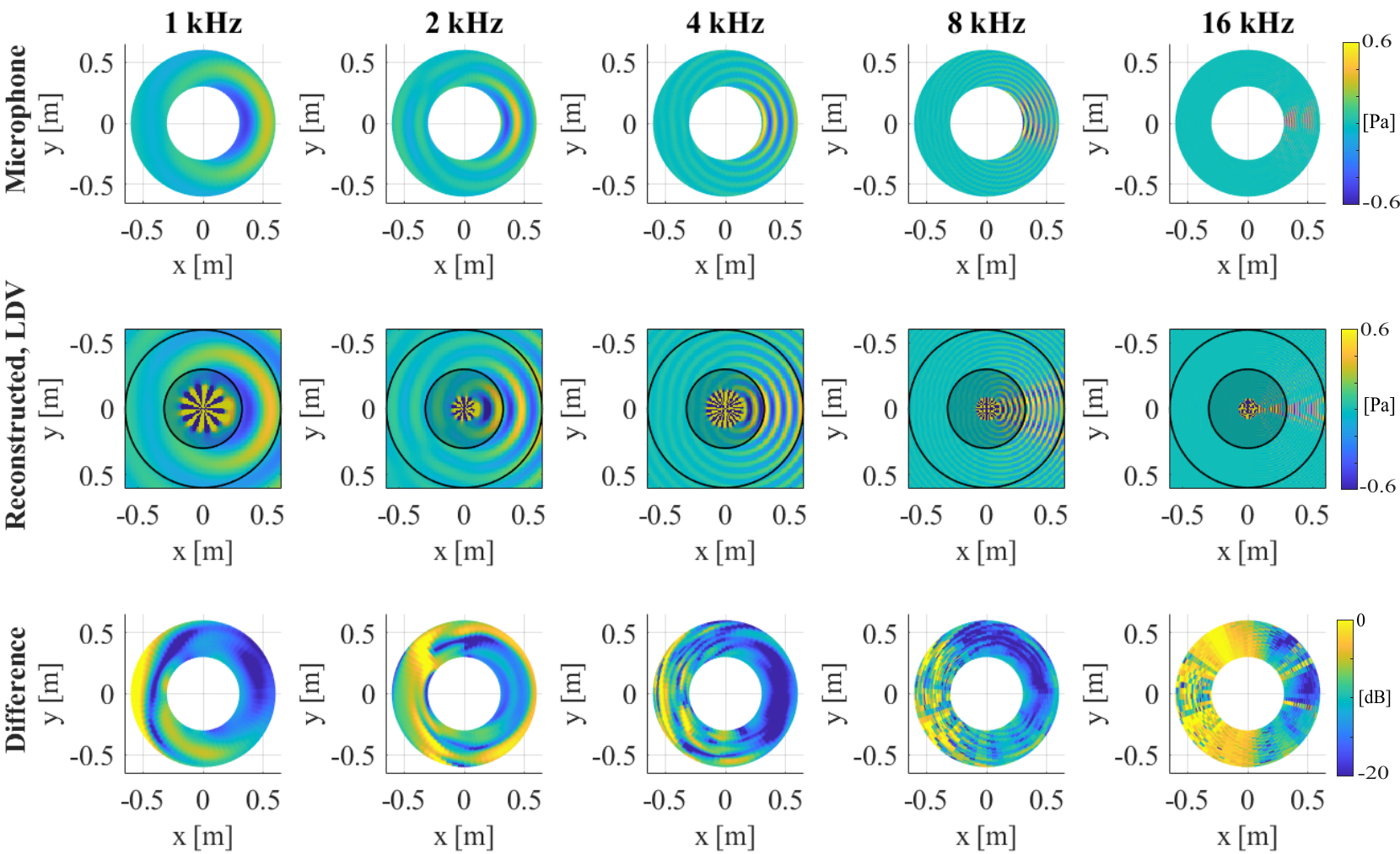}
\caption{Experimental sound field reconstruction results from LDV projections. Sound fields are created at different frequencies from 1$\,$kHz to 16$\,$kHz. The reference sound pressure is recorded by the microphone placed nearest the laser path.}
\label{Fig_9}
\end{figure*}

To validate the effectiveness of the proposed method for experimental data, we generated harmonic sound fields at frequencies 1$\,$kHz, 2$\,$kHz, 4$\,$kHz, 8$\,$kHz, and 16$\,$kHz and compared the reconstructed sound fields by CHE with the fields measured by a microphone.

Since we used a commercially available LDV with a displacement decoder that outputs an electric signal proportional to the displacement, 
the projections $s_m$ were obtained as follows:

\begin{itemize}
  \item The displacement signal of the LDV is proportional to the line-integral of sound pressure in our measurement, as explained in \cite{torras2012sound}.
  \item After digitizing the signal, we multiplied a coefficient to convert the voltage signal with the unit of $V$ to the line-integral of sound pressure having the unit of Pa$\cdot$ m.
  \item Then, the time-domain signal is Fourier transformed, and the complex amplitude at a given frequency $\omega$ is extracted. This complex value is the projection $s_m$.
\end{itemize}


The results are shown in Fig.~\ref{Fig_9}. It indicates that the proposed CHE algorithm effectively reconstructs the sound fields from LDV projections. It should be noted that our LDV device introduces a signal latency of 900$\,\mu$s, which necessitates compensation when comparing the reconstructed sound fields with the reference sound fields recorded by the microphone. 
The third row of Fig.~\ref{Fig_9} shows the differences between microphone and LDV reconstructions, which were calculated using the same method as estimating reconstruction errors, as discussed in Section \ref{simulatedreconstruction}.
The experimental results indicate variations in the discrepancy between the reconstructed LDV sound field and the microphone signal across different frequencies. For 1 to 8 kHz, the reconstruction results of the proposed method align well with the measurements obtained by the microphones, especially in front of the loudspeaker (positive $x$ direction). For the 16\,kHz sound field, however, the difference is quite significant, specifically showing that the discrepancies are around 0\,dB at most locations. 
This may indicate that the origin of the proposed method significantly impacts the reconstruction accuracy for high frequencies. 

\begin{figure*}[!t]
\centering
\includegraphics[width=0.9\textwidth]{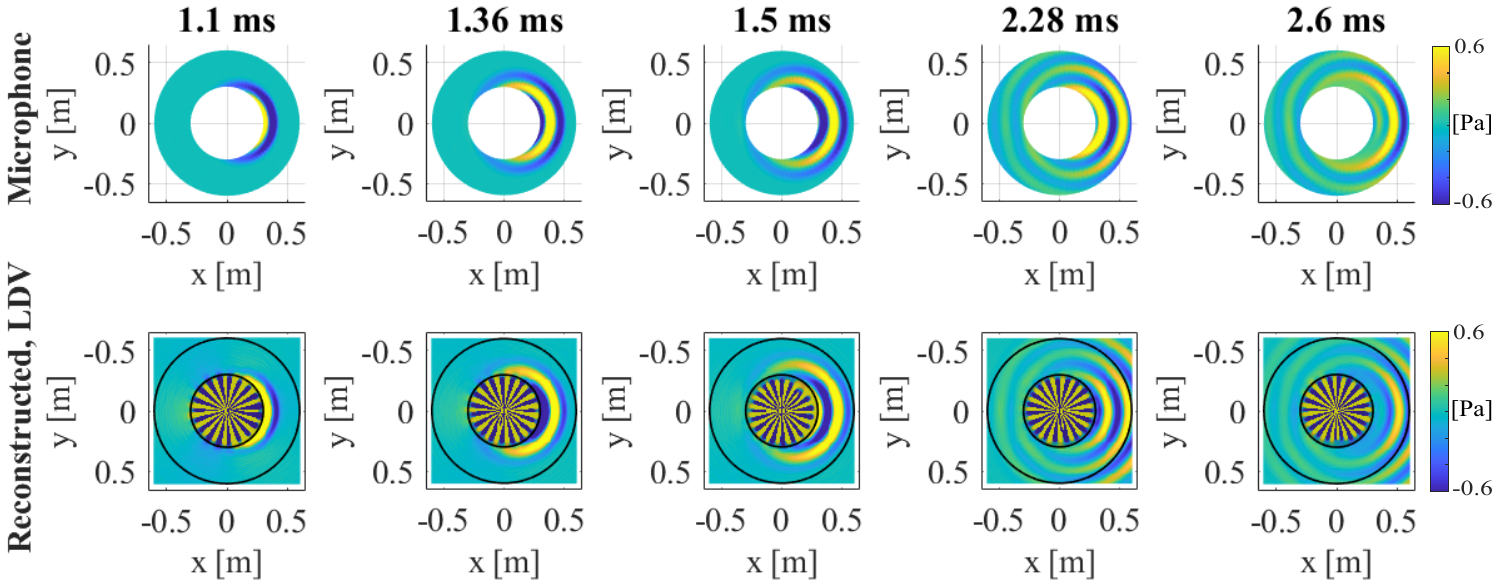}
\caption{Time-domain sound field reconstruction.}
\label{Fig_10}
\end{figure*}

\subsection{Time-domain sound-field reconstruction}
In order to assess the performance of time-domain sound field reconstruction, a 2$\,$kHz sinusoidal burst wave with a duration of 10$\,$ms was emitted from the loudspeaker. The experimental configuration utilized for this evaluation is consistent with the parameters defined in Section \ref{experimentsetup}. To reconstruct the sound field in the time domain from experimentally acquired data, the procedure comprises the following steps:

\begin{itemize}
  \item Step 1: Perform a Fast Fourier Transform (FFT) on all time-domain LDV projection data to convert projections from the time domain to the frequency domain.
  \item Step 2: Apply the proposed CHE algorithm to reconstruct the sound field in the frequency domain from the FFT output data obtained in Step 1, independently for each frequency.
  \item Step 3: Combine the 2-dimensional (2D) reconstructed sound fields at all frequencies into a 3-dimensional (3D) array. The third dimension represents frequencies ranging from 0$\,$Hz to the maximum frequency, with a length identical to the FFT length in Step 1.
  \item Step 4: Perform an inverse FFT on the 3D array of reconstructed sound fields in the frequency domain to yield the results of the time-domain sound field reconstruction.
\end{itemize}

Fig.~\ref{Fig_10} illustrates multiple frames extracted from time-domain sound fields reconstructed using LDV projections at various time points: 1.1$\,$ms, 1.36$\,$ms, 1.5$\,$ms, 2.28$\,$ms, and 2.6$\,$ms. Concurrently, time-domain microphone signals were acquired synchronously with the LDV device for reference. To enable a meaningful comparison with the microphone signals, a delay of 900$\,\mu$s in the LDV signal was corrected. Additionally, we empirically determined the response delay of the microphone system as approximately 10$\,\mu$s through experimental measurements. Upon scrutinizing the data, it is evident that minor temporal disparities of approximately 90$\,\mu$s persist between the sound fields captured by the microphone and those recorded by the LDV device. These slight variations may be attributed to undetermined signal delays within the LDV or microphone circuitry. Nevertheless, it is noteworthy that the sound fields reconstructed using the CHE algorithm from LDV projections exhibit a remarkable similarity to the corresponding microphone signals.

\section{Discussions}
In order to expedite the process of sound field reconstruction, two key aspects are considered: determining an appropriate number of circular harmonics $N$ and reducing the number of required projections $M$. 

\subsection{Influence of expansion order}

 \begin{figure}[!t]
\centering
\includegraphics[width=0.49\textwidth]{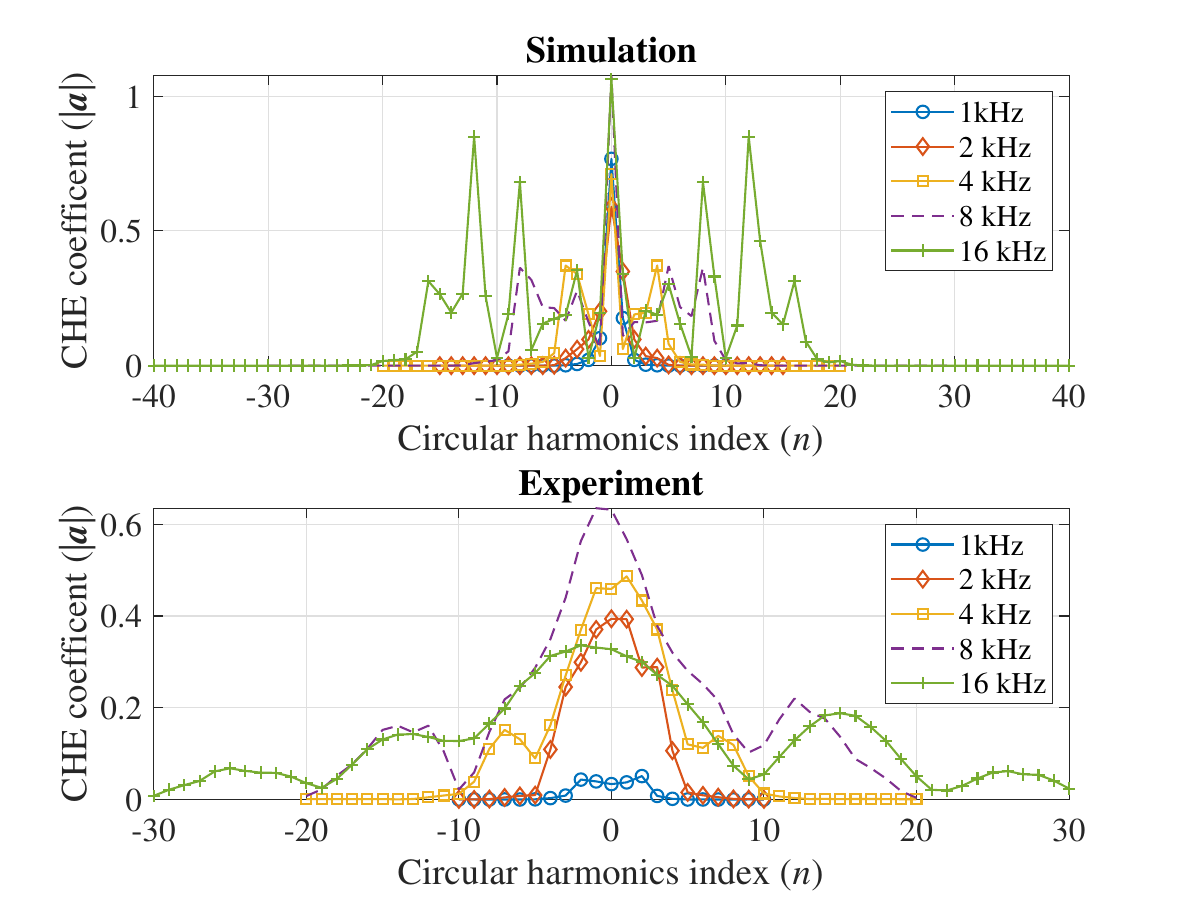}
\caption{Determine the necessary count of circular harmonic waves essential for a complete representation of the amplitude of the CHE coefficient ($\left| \mathbf{a} \right|$) in sound fields across a range of frequencies, specifically, at 1$\,$kHz, 2$\,$kHz, 4$\,$kHz, 8$\,$kHz, and 16$\,$kHz. This assessment is performed in conjunction with the experiments shown in Fig.~\ref{Fig_9} and simulations involving off-center sound sources at coordinates identical to those used in our experiments.}
\label{Fig_11}
\end{figure}

The appropriate selection of the value $N$ is visually depicted in Fig.~\ref{Fig_11}. This figure illustrates the number of circular harmonic waves required to fully represent the amplitude of the expansion coefficients, denoted as $\left| \mathbf{a} \right|$. We have investigated both experimentally and by simulation to come up with a suitable value range for the number of circular harmonics waves $N$, because the larger the number of circular harmonics waves, the higher the complexity of the sound reconstruction algorithm, leading to a longer time to reconstruct the sound field. When we examine our experimental data, we determine that the suitable $N$ values are 5, 10, 15, 20, and 30 for sound fields at 1$\,$kHz, 2$\,$kHz, 4$\,$kHz, 8$\,$kHz, and 16$\,$kHz, respectively. This selection is based on the fact that the amplitude of expansion coefficients approaches zero or nearly zero when the index of the circular harmonic $n$ is in the range [$-N$:$\,$$N$]. Besides, computational simulations also suggest that the appropriate $N$ values are 5, 10, 15, 20, and 30 for sound fields at 1$\,$kHz, 2$\,$kHz, 4$\,$kHz, 8$\,$kHz, and 16$\,$kHz, respectively, which are similar to the results obtained from experiments. It is worth noting that higher frequencies necessitate larger values for $N$, which in turn require more computational time for the reconstruction process. Therefore, in our proposed CHE method, we opted for $N$ values of 10 for 1$\,$kHz, 15 for 2$\,$kHz, 20 for 4$\,$kHz, 30 for 8$\,$kHz, and 40 for 16$\,$kHz, as these values fully represent the CHE coefficient amplitudes. 

\subsection{Influence of number of projections}
\begin{figure}[!t]
\centering
\includegraphics[width=0.5\textwidth]{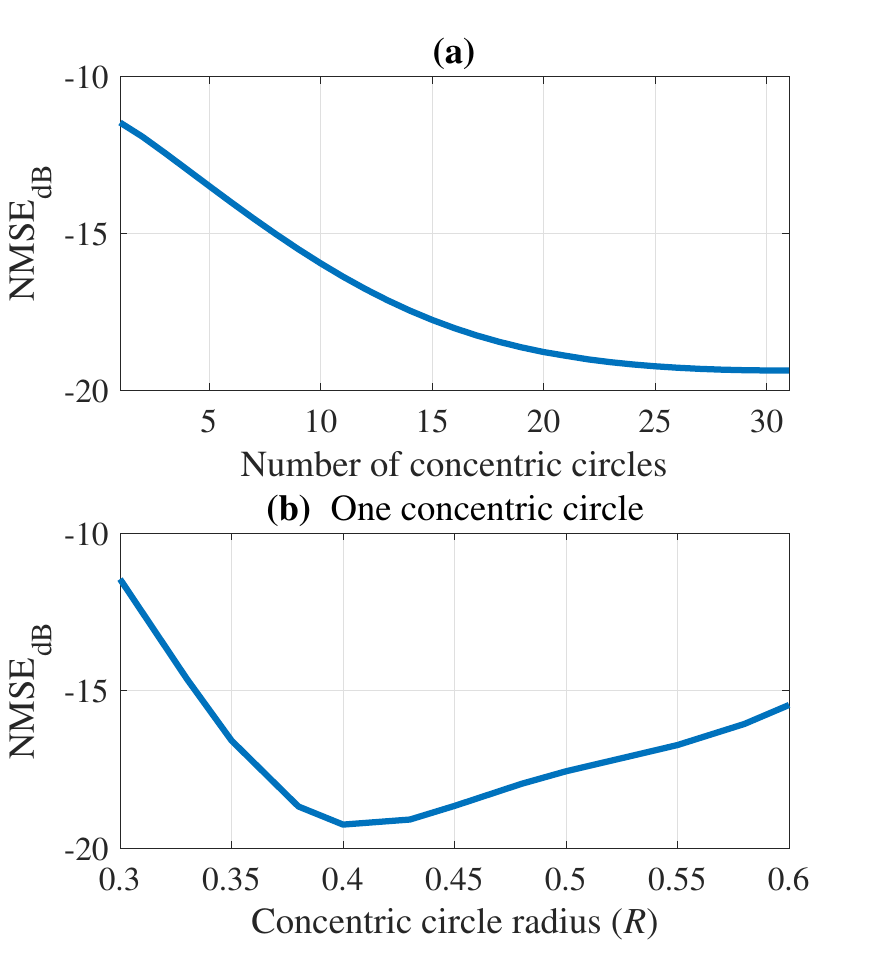}
\caption{The assessment of the NMSE in decibels (dB) for 2$\,$kHz reconstructed sound fields involves two scenarios: \textbf{(a)} the NMSE is evaluated by varying the number of projections acquired from concentric circles with radii ranging from $R_\mathrm{min}\,$=$\,0.3\,$m to $R_\mathrm{max}\,$=$\,0.6\,$m; \textbf{(b)} the NMSE is assessed using 72 projections from a single concentric circle.}
\label{Fig_12}
\end{figure}

In order to determine the optimal number of projections necessary for sound field reconstruction, we conducted an assessment of the reconstructed sound field quality employing NMSE analysis, $\mathrm{NMS{E_{dB}}} = 20{\log _{10}}({{{{\left\| {\tilde p -  p} \right\|}_2}} \mathord{\left/
{\vphantom {{{{\left\| {\tilde p - p} \right\|}_2}} {{{\left\| p \right\|}_2}}}} \right.
\kern-\nulldelimiterspace} {{{\left\| p \right\|}_2}}})$, where $\tilde p$ and $p$ indicate reconstructed sound field and sound field recorded from microphone, respectively.
For reference sound field, an experimental 2$\,$kHz sound signal was created from a loudspeaker placed in an anechoic chamber, with the sound source positioned at an off-center location (0.12, 0). In this analysis, we examined a range of projection quantities, from 72 projections originating from the innermost concentric circle with a radius of $0.3\,$m, to 2232 projections obtained from 31 concentric circles, with radii varying from $0.3\,$m to $0.6\,$m. Fig.\,\ref{Fig_12}(a) shows the NMSEs for different numbers of concentric circles. It can be seen that the highest quality is achieved with 31 concentric circles, yielding an NMSE of -19.4\,dB. Moreover, Fig.~\ref{Fig_12}(a) demonstrates that the NMSE nearly reaches saturation at -19.2\,dB when the number of concentric circles exceeds 25.

To minimize the number of projections employed in the reconstruction process, it is essential to identify concentric circles that yield optimal reconstruction performance among all available concentric circles. Fig.~\ref{Fig_12}(b) plots the NMSEs reconstructed from single concentric circle measurement. It demonstrates that the choice of concentric circle radius has a significant impact on reconstruction quality. Notably, $R\,$=$\,0.4\,$m yields superior results, with the NMSE of -19.2$\,$dB, surpassing other radii.
It would be surprising that the reconstruction error from the single circle of $R = 0.4\,$m is comparable to those from 25 to 31 circles as can be seen in Fig.~\ref{Fig_12}(a). This indicates that the proposed CHE method, combined with the appropriate selection of circle radii, can significantly reduce the number of projections required for sound field reconstruction, thereby significantly reducing the complexity of implementing the method as well as the processing time of the algorithm. The determination of the optimal concentric circle radius and the appropriate quantity of concentric circles is beyond the scope of this paper. However, we envision that our discoveries may provide a foundational basis for future research endeavors aimed at investigating these intriguing inquiries.

 \begin{figure}[!t]
\centering
\includegraphics[width=0.5\textwidth]{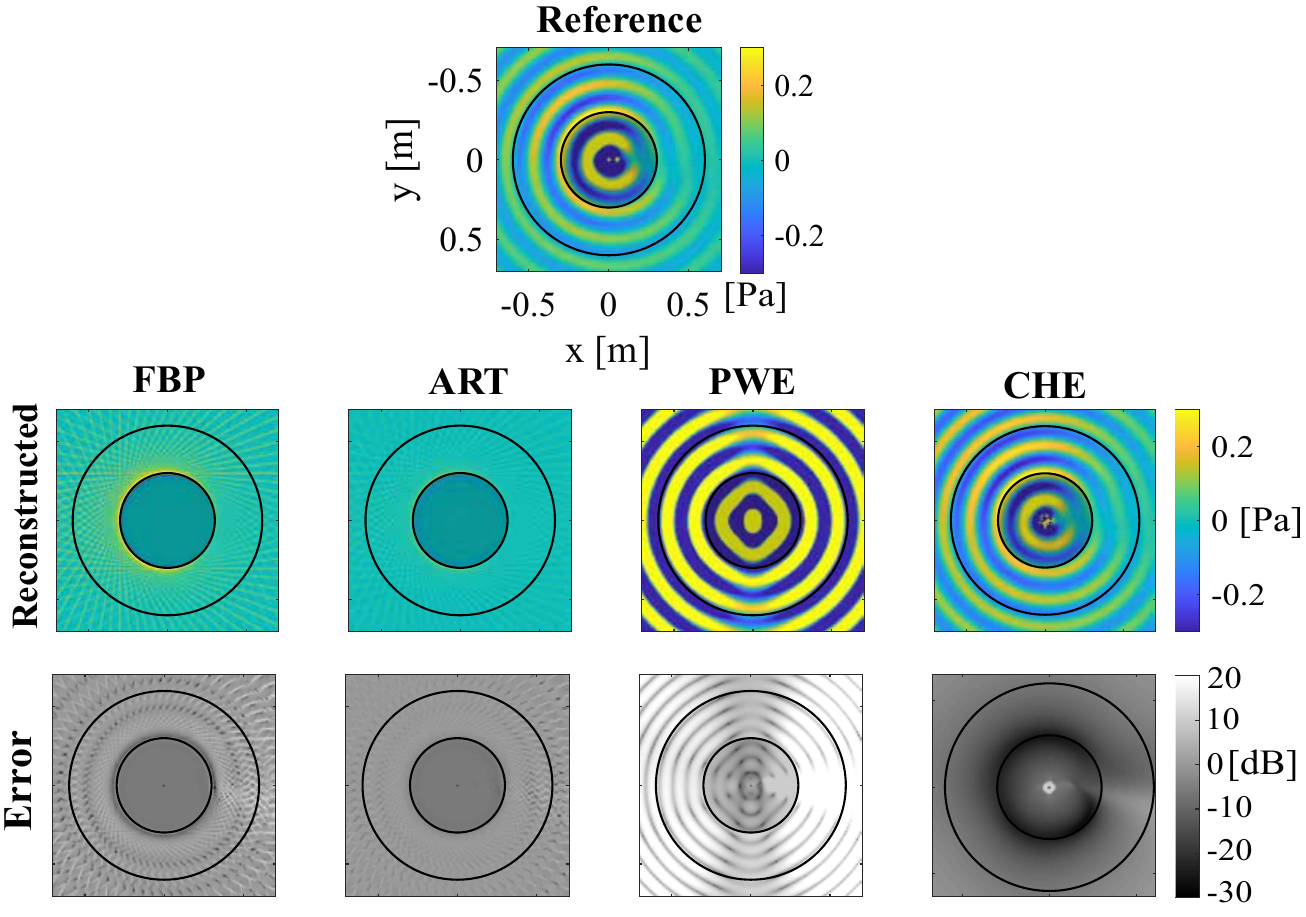}
\caption{Comparison of sound field reconstruction quality at 2\,kHz using different algorithms with the same number of projections, which is 72. For the CHE algorithm, 72 projections are taken from a concentric circle with a radius of $R=0.3\,$m.}
\label{Fig_13}
\end{figure}

Figure~\ref{Fig_13} presents the comparison results of sound field reconstruction at a frequency of 2 kHz using the algorithms FBP, ART, PWE, and CHE from the 72 projections at $R = 0.3\,$m. Apparently, FBP, ART, and PWE algorithms fail to reconstruct the original sound field,  
It is noteworthy that when applying the PWE algorithm to reconstruct the sound field, there is a significant difference in the amplitude of the reconstructed sound field compared to the original sound field, leading to a huge error.
In contrast, the proposed CHE algorithm demonstrates successful sound field reconstruction with only 72 projections. 
Specifically, the error image shows that the errors of the CHE method are in the range of -30\,dB to -10\,dB.

\section{Conclusion}
This paper introduces an innovative approach involving concentric circle sampling and a two-dimensional exterior sound-field reconstruction technique based on the extension of circular harmonics. The proposed reconstruction method effectively reconstructs the sound field, particularly when the sound source is located within the reconstruction area. To evaluate the effectiveness of the proposed approach, comprehensive assessments are conducted through numerical simulations and real-world experiments. The outcomes of these simulations and experiments compellingly establish that the proposed technique outperforms conventional reconstruction methods, exhibiting superior accuracy, and concurrently minimizing the number of required measured projections for the reconstruction process. This study introduces significant advancements in acoustic data measurement and analysis, allowing for comprehensive spatiotemporal sound field capture across extensive spatial regions. Moreover, this methodology holds potential value in diverse disciplines that focus on wave field measurement and visualization. Some potential applications of this research include loudspeaker characterization, noise pattern analysis, and so forth.

 \section*{Acknowledgment}
The authors thank Dr. Samuel A. Verburg for his helpful comments on the plane wave expansion reconstruction method in early discussions. We extend our sincere gratitude to the anonymous reviewers whose invaluable feedback and constructive critiques significantly contributed to enhancing the overall quality and rigor of this paper.

\begin{appendices}
\section{Interior and Exterior Problems}

The distinction between interior and exterior problems is a natural consequence of theoretical acoustics~\cite{williams2000fourier}.  
Let the Helmholtz equation in polar coordinates be given by  
\begin{equation}
    (\nabla^2 + k^2) p(r, \phi) = 0,
\end{equation}
where  
\begin{equation}
    \nabla^2 = \frac{\partial^2}{\partial r^2} + \frac{1}{r} \frac{\partial}{\partial r} + \frac{1}{r^2} \frac{\partial^2}{\partial \phi^2}
\end{equation}
is the Laplacian operator in polar coordinates.  
General solutions to this equation can be expressed in the following two equivalent forms:
\begin{eqnarray}
p(r, \phi) = \!\! \sum\limits_{n = -\infty }^\infty \left[ c_n^{(1)}(\omega) J_n(kr) + c_n^{(2)}(\omega) Y_n(kr) \right] e^{jn\phi} \label{eq:ab} \\
\!\! = \!\! \sum\limits_{n = -\infty }^\infty \left[ a_n^{(1)}(\omega) H_n^{(1)}(kr) + a_n^{(2)}(\omega) H_n^{(2)}(kr) \right] e^{jn\phi}, \label{eq:cd}
\end{eqnarray}
where $J_n$ and $Y_n$ are the Bessel functions of the first and second kinds, and $H_n^{(1)}$ and $H_n^{(2)}$ are the Hankel functions of the first and second kinds, respectively. The coefficients $c_n^{(1)}, c_n^{(2)}, a_n^{(1)}, a_n^{(2)}$ are the corresponding expansion coefficients.  
Both Eq.~\eqref{eq:ab} and Eq.~\eqref{eq:cd} provide complete and equivalent solutions to the Helmholtz equation with infinite-order summations. However, since each of the four basis functions has a distinct physical meaning, the choice of solution form is crucial when truncating the series for numerical computation.

In the interior problem, where the sound source lies outside the region of interest (as illustrated in Fig.~\ref{Fig_1}), the pressure field must remain finite at the expansion origin. Since $Y_n$ and both Hankel functions diverge at the origin, the interior field can be expressed solely using $J_n$ as:
\begin{equation}
    p_\text{int} (r, \phi) = \sum\limits_{n = -\infty }^\infty c_n^{(1)}(\omega) J_n(kr) e^{jn\phi}.
\end{equation}

In contrast, the exterior problem considers all sound sources to be enclosed within a finite region around the origin. In this case, the sound field must represent an outward-propagating wave from the origin to infinity. Among the four basis functions, only $H_n^{(2)}$ satisfies this condition\footnote{Note that the propagation direction of $H_n^{(1)}$ and $H_n^{(2)}$ depends on the sign of the time-dependence term. We employ $\mathrm{e}^{i \omega t}$ for time convention, under which $H_n^{(2)}$ represents outgoing waves.}. Thus, the solution to the exterior problem is given by:
\begin{equation}
    p_\text{ext} (r, \phi) = \sum\limits_{n = -\infty }^\infty a_n^{(2)}(\omega) H_n^{(2)}(kr) e^{jn\phi}.
\end{equation}
In our work, we adopt this representation and omit the superscript for simplicity.  
For a complete derivation and interpretation, please refer to the textbook~\cite{williams2000fourier}.

It should be noted that existing PWE methods are derived based on the interior solution of the Helmholtz equation in Cartesian coordinates. As a result, they are not valid for exterior problems. If the measured projections surround a sound source, PWE leads to poor reconstruction performance, as demonstrated throughout this paper.

\end{appendices}
\color{black}
	
\bibliographystyle{IEEEtran}

\begin{thebibliography}{10}
\providecommand{\url}[1]{#1}
\csname url@samestyle\endcsname
\providecommand{\newblock}{\relax}
\providecommand{\bibinfo}[2]{#2}
\providecommand{\BIBentrySTDinterwordspacing}{\spaceskip=0pt\relax}
\providecommand{\BIBentryALTinterwordstretchfactor}{4}
\providecommand{\BIBentryALTinterwordspacing}{\spaceskip=\fontdimen2\font plus
\BIBentryALTinterwordstretchfactor\fontdimen3\font minus \fontdimen4\font\relax}
\providecommand{\BIBforeignlanguage}[2]{{%
\expandafter\ifx\csname l@#1\endcsname\relax
\typeout{** WARNING: IEEEtran.bst: No hyphenation pattern has been}%
\typeout{** loaded for the language `#1'. Using the pattern for}%
\typeout{** the default language instead.}%
\else
\language=\csname l@#1\endcsname
\fi
#2}}
\providecommand{\BIBdecl}{\relax}
\BIBdecl

\bibitem{torras2012sound}
A.~Torras-Rosell, S.~Barrera-Figueroa, and F.~Jacobsen, ``Sound field reconstruction using acousto-optic tomography,'' \emph{The Journal of the Acoustical Society of America}, vol. 131, no.~5, pp. 3786--3793, 2012.

\bibitem{AcousticsToday}
S.~A. Verburg, K.~Ishikawa, E.~Fernandez-Grande, and Y.~Oikawa, ``A century of acousto-optics- from early discoveries to modern sensing of sound with light,'' \emph{Acoust. Today}, vol.~19, pp. 54--62, 2023.

\bibitem{Nakamura2002}
K.~Nakamura, M.~Hirayama, and S.~Ueha, ``Measurements of air-borne ultrasound by detecting the modulation in optical refractive index of air,'' in \emph{Proc. IEEE Ultrasonics Symposium}, 2002, pp. 609--612.

\bibitem{Harland2002}
A.~Harland, J.~Petzing, and J.~Tyrer, ``Non-invasive measurements of underwater pressure fields using laser doppler velocimetry,'' \emph{J. Sound Vib.}, vol. 252, pp. 169--177, 4 2002.

\bibitem{Ishikawa2016}
K.~Ishikawa, K.~Yatabe, N.~Chitanont, Y.~Ikeda, Y.~Oikawa, T.~Onuma, H.~Niwa, and M.~Yoshii, ``High-speed imaging of sound using parallel phase-shifting interferometry,'' \emph{Opt. Express}, vol.~24, p. 12922, 6 2016.

\bibitem{Matoba2014}
O.~Matoba, H.~Inokuchi, K.~Nitta, and Y.~Awatsuji, ``Optical voice recorder by off-axis digital holography,'' \emph{Opt. Lett.}, vol.~39, p. 6549, 11 2014.

\bibitem{Ishikawa2018}
K.~Ishikawa, R.~Tanigawa, K.~Yatabe, Y.~Oikawa, T.~Onuma, and H.~Niwa, ``Simultaneous imaging of flow and sound using high-speed parallel phase-shifting interferometry,'' \emph{Opt. Lett.}, vol.~43, p. 991, 3 2018.

\bibitem{Rajput2019}
S.~K. Rajput, O.~Matoba, and Y.~Awatsuji, ``Holographic multi-parameter imaging of dynamic phenomena with visual and audio features,'' \emph{Opt. Lett.}, vol.~44, p. 995, 2 2019.

\bibitem{Takase2021}
Y.~Takase, K.~Shimizu, S.~Mochida, T.~Inoue, K.~Nishio, S.~K. Rajput, O.~Matoba, P.~Xia, and Y.~Awatsuji, ``High-speed imaging of the sound field by parallel phase-shifting digital holography,'' \emph{Appl. Opt.}, vol.~60, p. A179, 2 2021.

\bibitem{Rajput2021}
S.~K. Rajput, O.~Matoba, M.~Kumar, X.~Quan, and Y.~Awatsuji, ``Sound wave detection by common-path digital holography,'' \emph{Opt. Lasers Eng.}, vol. 137, p. 106331, 2 2021.

\bibitem{Zhong2022}
Z.~Zhong, C.~Wang, L.~Liu, Y.~Liu, L.~Yu, B.~Liu, and M.~Shan, ``Visual measurement of instable sound field using common-path off-axis digital holography,'' \emph{Opt. Lasers Eng.}, vol. 158, p. 107129, 11 2022.

\bibitem{Ishikawa2021}
K.~Ishikawa, Y.~Shiraki, T.~Moriya, A.~Ishizawa, K.~Hitachi, and K.~Oguri, ``Low-noise optical measurement of sound using midfringe locked interferometer with differential detection,'' \emph{J. Acoust. Soc. Am.}, vol. 150, pp. 1514--1523, 8 2021.

\bibitem{flannery1987three}
B.~P. Flannery, H.~W. Deckman, W.~G. Roberge, and K.~L. D'Amico, ``Three-dimensional x-ray microtomography,'' \emph{Science}, vol. 237, no. 4821, pp. 1439--1444, 1987.

\bibitem{Koyama2024}
S.~Koyama, J.~G.~C. Ribeiro, T.~Nakamura, N.~Ueno, and M.~Pezzoli, ``Physics-informed machine learning for sound field estimation: Fundamentals, state of the art, and challenges,'' \emph{IEEE Signal Processing Magazine}, vol.~41, no.~6, pp. 60--71, 2024.

\bibitem{Ochmann1999}
M.~Ochmann, ``The full-field equations for acoustic radiation and scattering,'' \emph{The Journal of the Acoustical Society of America}, vol. 105, no.~5, pp. 2574--2584, 05 1999.

\bibitem{williams2000fourier}
E.~G. Williams and J.~A. Mann~III, ``Fourier acoustics: sound radiation and nearfield acoustical holography,'' 2000.

\bibitem{yatabe2017acousto}
K.~Yatabe, K.~Ishikawa, and Y.~Oikawa, ``Acousto-optic back-projection: Physical-model-based sound field reconstruction from optical projections,'' \emph{Journal of Sound and Vibration}, vol. 394, pp. 171--184, 2017.

\bibitem{verburg2021acousto}
S.~A. Verburg and E.~Fernandez-Grande, ``Acousto-optical volumetric sensing of acoustic fields,'' \emph{Physical Review Applied}, vol.~16, no.~4, p. 044033, 2021.

\bibitem{ishikawa2021physical}
K.~Ishikawa, K.~Yatabe, and Y.~Oikawa, ``Physical-model-based reconstruction of axisymmetric three-dimensional sound field from optical interferometric measurement,'' \emph{Measurement Science and Technology}, vol.~32, no.~4, p. 045202, 2021.

\bibitem{Verburg2022}
S.~A. Verburg, E.~G. Williams, and E.~Fernandez-Grande, ``Acousto-optic holography,'' \emph{J. Acoust. Soc. Am.}, vol. 152, pp. 3790--3799, 12 2022.

\bibitem{kak2001principles}
A.~C. Kak and M.~Slaney, \emph{Principles of computerized tomographic imaging}.\hskip 1em plus 0.5em minus 0.4em\relax SIAM, 2001.

\bibitem{Oikawa2005}
Y.~Oikawa, Y.~Ikeda, M.~Goto, T.~Takizawa, and Y.~Yamasaki, ``Sound field measurements based on reconstruction from laser projections,'' in \emph{Proc. (ICASSP '05). IEEE International Conference on Acoustics, Speech, and Signal Processing, 2005.}, vol.~4.\hskip 1em plus 0.5em minus 0.4em\relax IEEE, pp. 661--664.

\bibitem{koponen2019acoustic}
E.~Koponen, J.~Leskinen, T.~Tarvainen, and A.~Pulkkinen, ``Acoustic pressure field estimation methods for synthetic schlieren tomography,'' \emph{The Journal of the Acoustical Society of America}, vol. 145, no.~4, pp. 2470--2479, 2019.

\bibitem{HERMAWANTO2020}
D.~Hermawanto, K.~Ishikawa, K.~Yatabe, and Y.~Oikawa, ``Determination of frequency response of mems microphone from sound field measurements using optical phase-shifting interferometry method,'' \emph{Applied Acoustics}, vol. 170, p. 107523, 2020.

\bibitem{andersen1984simultaneous}
A.~H. Andersen and A.~C. Kak, ``Simultaneous algebraic reconstruction technique (sart): a superior implementation of the art algorithm,'' \emph{Ultrasonic imaging}, vol.~6, no.~1, pp. 81--94, 1984.

\bibitem{herman2009fundamentals}
G.~T. Herman, \emph{Fundamentals of computerized tomography: image reconstruction from projections}.\hskip 1em plus 0.5em minus 0.4em\relax Springer Science \& Business Media, 2009.

\bibitem{natterer2001mathematics}
F.~Natterer, \emph{The mathematics of computerized tomography}.\hskip 1em plus 0.5em minus 0.4em\relax SIAM, 2001.

\bibitem{wang2015optical}
Q.~Wang, Q.~Guo, L.~Lei, and J.~Zhou, ``Optical interference-based image encryption using circular harmonic expansion and spherical illumination in gyrator transform domain,'' \emph{Optics Communications}, vol. 346, pp. 124--132, 2015.

\bibitem{arsenault1986properties}
H.~H. Arsenault and Y.~Sheng, ``Properties of the circular harmonic expansion for rotation-invariant pattern recognition,'' \emph{Applied Optics}, vol.~25, no.~18, pp. 3225--3229, 1986.

\bibitem{ren2020two}
Y.~Ren and Y.~Haneda, ``Two-dimensional exterior sound field reproduction using two rigid circular loudspeaker arrays,'' \emph{The Journal of the Acoustical Society of America}, vol. 148, no.~4, pp. 2236--2247, 2020.

\end{thebibliography}


\begin{IEEEbiography}[{\includegraphics[width=1in,height=1.25in,clip,keepaspectratio]{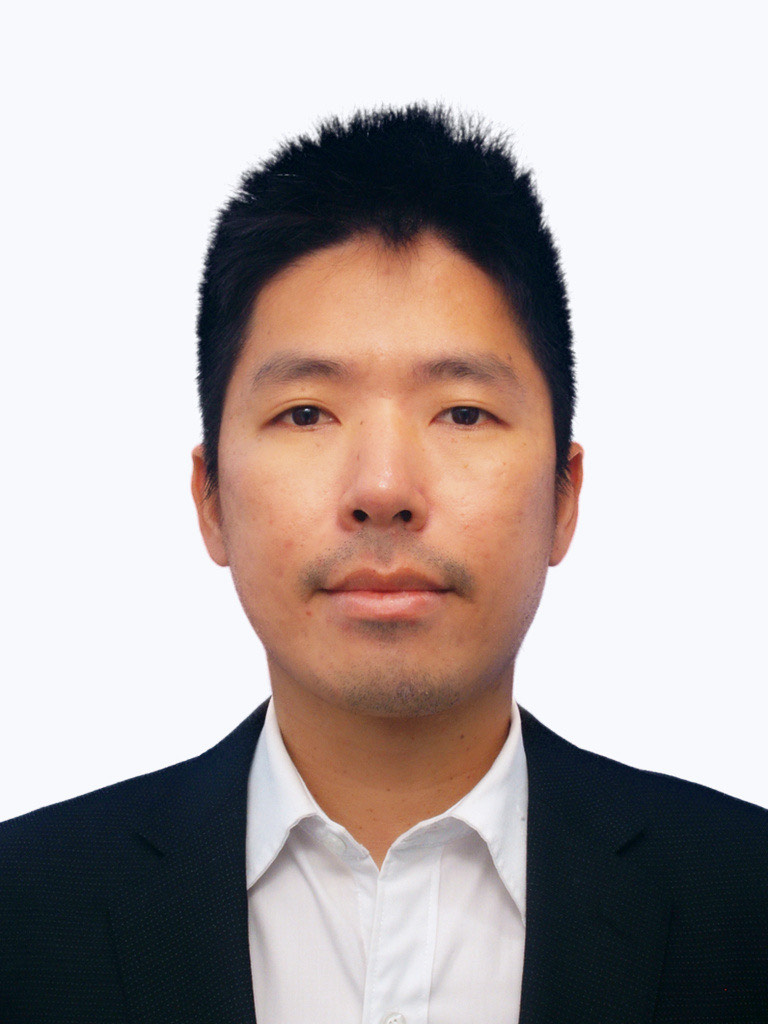}}]{Phuc Duc Nguyen} (Member, IEEE) received B.S. and M.S. degrees in Electronics and Telecommunications in 2012, and 2015, both from the Vietnam National University (VNU-HCM). He received his Ph.D. degree in information science at the Nara Institute of Science and Technology (NAIST), Japan, in 2018. Phuc has been a lecturer at the Posts and Telecommunication Institute of Technology, Vietnam, from 2018 to 2020. He was a visiting postdoctoral scholar at École Nationale Supérieure de l'Electronique et de ses Applications, Cergy-Paris, France, in 2019. In the Fall of 2020, Phuc joined the NTT Communication Science Laboratories in Japan as a research associate. Phuc is a recipient of the Best Paper award in ATC'18 and the Best Oral Presentation award in ICFCC'17. His research interests include optical wireless communications, optical information processing, and FPGA-based hardware accelerators. He is a member of IEEE, IEEE Photonics Society, IEICE, and the Acoustical Society of Japan (ASJ).
	\end{IEEEbiography}
	
	\begin{IEEEbiography}[{\includegraphics[width=1in,height=1.25in,clip,keepaspectratio]{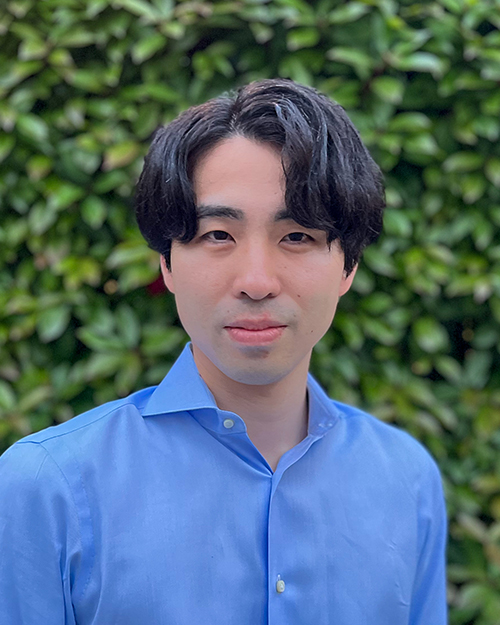}}]{Kenji Ishikawa} (Member, IEEE) received the M.S. and Ph.D. degrees in engineering from Dept. Intermedia Art and Science, Waseda University, in 2015 and 2019. In 2019, he joined NTT Communication Science Laboratories as a researcher. His research interests lie in the interdisciplinary area of acoustics, optics, and measurement technologies. Dr. Ishikawa is a recipient of the 42nd Awaya Kiyoshi Award (2017) and Itakura Prize Innovative Young Researcher Award (2023) from the Acoustical Society of Japan. He is a member of IEEE, the Acoustical Society of Japan (ASJ) and the Laser Society of Japan (LSJ).
	\end{IEEEbiography}	
	
	\begin{IEEEbiography}[{\includegraphics[width=1in,height=1.25in,clip,keepaspectratio]{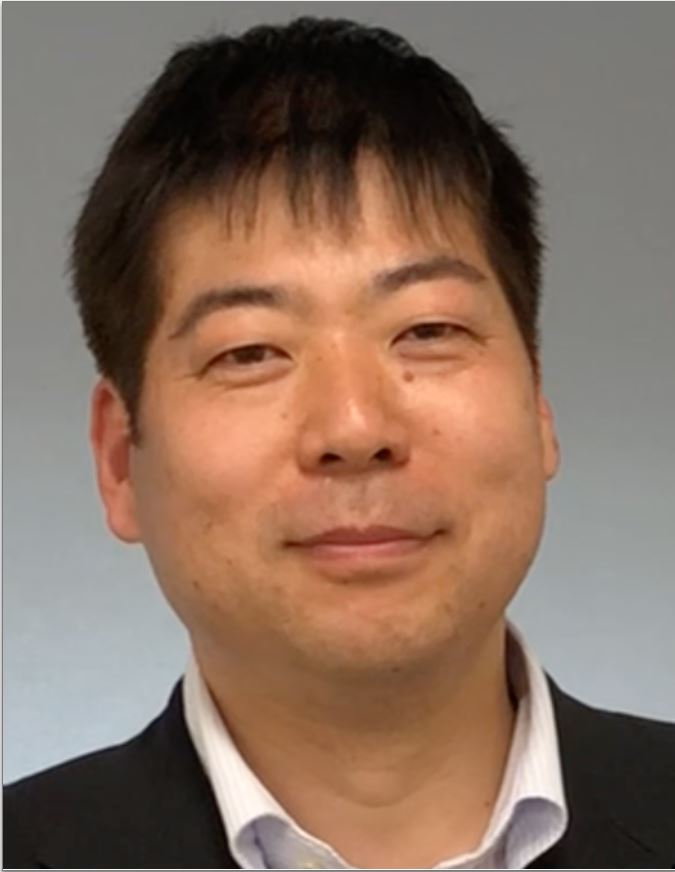}}]{Noboru Harada} (Senior Member, IEEE) received the B.S. and M.S. degrees in computer science and systems engineering from the Kyushu Institute of Technology, Fukuoka, Japan, in 1995 and 1997, respectively, and the Ph.D. degree in computer science from the University of Tsukuba, Ibaraki, Japan, in 2017. Since joining NTT Corporation, Tokyo, Japan, in 1997, he has been involved with research on speech and audio signal processing, such as high-efficiency coding, lossless compression, and acoustic event detection, including anomaly sound detection.
	\end{IEEEbiography}
	
	\begin{IEEEbiography}[{\includegraphics[width=1in,height=1.25in,clip,keepaspectratio]{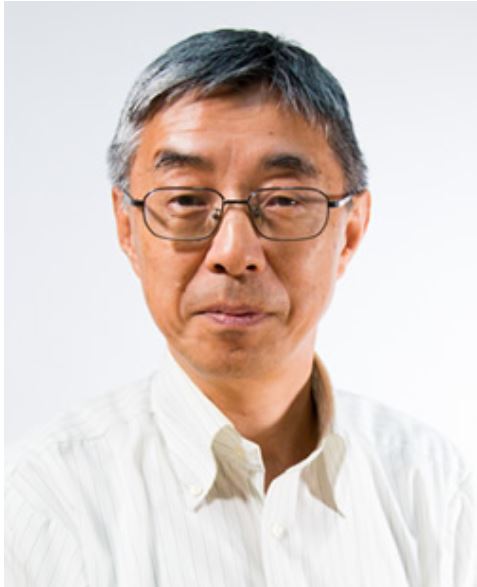}}]{Takehiro Moriya} (Life Fellow, IEEE) received the B.S., M.S., and Ph.D. degrees in mathematical engineering and instrumentation physics from the University of Tokyo, in 1978, 1980, and 1989, respectively. Since joining NTT laboratories in 1980, he has been engaged in research on medium to low bitrate speech and audio coding. In 1989, he worked with AT\&T Bell Laboratories as a Visiting Researcher. Since 1990, he has contributed to the standardization of coding schemes for the Japanese Public Digital Cellular system, ITU-T, ISO/IEC MPEG, and 3GPP. He is currently the Head of the Moriya Research Laboratory with NTT Communication Science Laboratories. Dr. Moriya is a recipient of many awards, including IEEE James L. Flanagan Speech and Audio Processing Award in 2016. He is currently an NTT Fellow. He is a member of IPSJ, Acoustic Engineering Society, and ASJ. He is also a member of IEEE Technical Committee on Speech and Language Processing.
	\end{IEEEbiography}	
\end{document}